\documentclass[aps,prl,twocolumn]{revtex4}
 \usepackage[usenames,dvipsnames]{color}
\usepackage{graphicx}
\usepackage{dcolumn}
\usepackage{bm}
\usepackage{amsmath}
\usepackage{gensymb}
\usepackage{amsfonts}
\usepackage{ragged2e}
\usepackage{psfrag}
\usepackage[pdftex,bookmarks,colorlinks=false]{hyperref}

\newcommand{\Eq}[1]{Eq. (\ref{#1})}
\newcommand{\vect}[1]{\boldsymbol{#1}}

\begin{document}
\title{Coherent coupling between localised and propagating phonon polaritons}

\author{Christopher R. Gubbin$^1$}
\author{Francesco Martini$^2$}
\author{Alberto Politi$^2$}
\author{Stefan A. Maier$^1$}
\author{Simone \surname{De Liberato}$^2$}

\affiliation{$^1$Blackett Laboratory, Imperial College London, London SW7 2AZ United Kingdom}
\affiliation{$^2$School of Physics and Astronomy, University of Southampton, Southampton, SO17 1BJ, United Kingdom}

\begin{abstract}
Following the recent observation of localised phonon polaritons  in user-defined silicon carbide nano-resonators, here we demonstrate coherent coupling between those localised modes and propagating phonon polaritons bound to the surface of the nano-resonator's substrate. In order to obtain phase-matching, the nano-resonators have been fabricated to serve the double function of hosting the localised modes, while also acting as grating for the propagating ones.
The coherent coupling between long lived, optically accessible localised modes, and low-loss propagative ones, opens the way to the design and realisation of phonon-polariton based quantum circuits.
\end{abstract}

\maketitle
\begin{figure*}
\includegraphics[width=\textwidth]{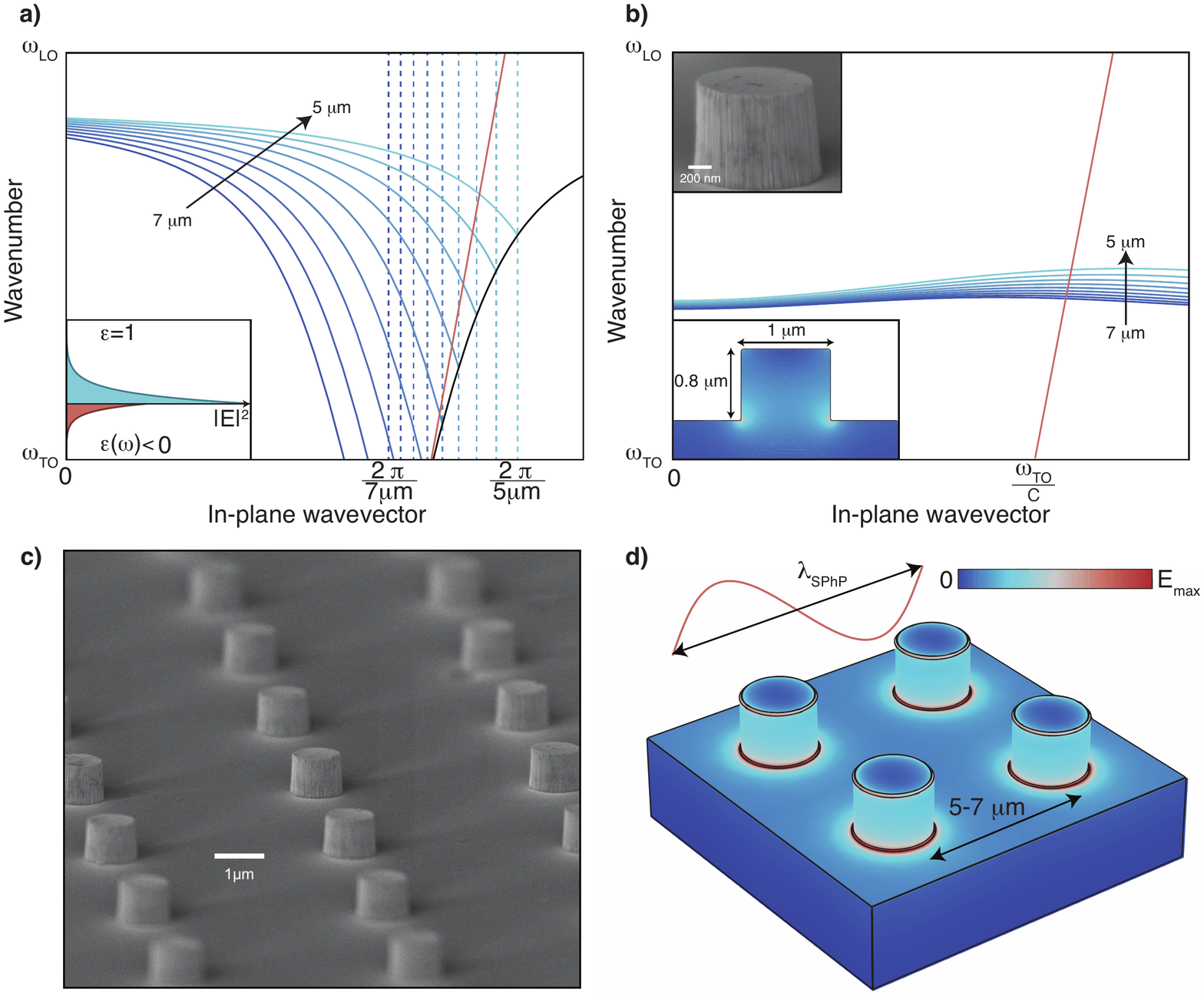} 
\centering
\caption{ \label{fig:Fig1} {\bf Surface and localised modes.} a) Fundamental dispersion of the surface mode is given by the black curve. Solid blue curves indicate surface mode folding from the edge of the first Brillouin zone (indicated by corresponding vertical dashed lines) for periodicities $5\mu$m-$7\mu$m. The red curve shows the vacuum light line. The inset shows the electric field norm for a surface mode at an air / SiC interface.
b) Tight-binding dispersion of the monopolar mode of a pillar array is indicated by blue curves for a variety of periodicities. The red curve is the vacuum light line, with $c$ the speed of light.
Insets show a slice of the mode electric field norm in an isolated SiC cylinder on substrate, calculated using COMSOL Multiphysics, and a SEM image of a fabricated resonator.
c) SEM image of the fabricated array
d) Electric field norm for a mode of the coupled array. The sinusoid indicates the surface mode wavelength.}
\end{figure*}

One of the basic requirements of electromagnetic quantum circuitry is a stable materials platform for coherent energy exchange between propagating and highly localised optical modes. Surface phonon polaritons are surface-bound, propagative modes arising from collective oscillations of ions at the surface of polar crystals, analogous to surface plasmon polaritons on metallic surfaces \cite{Borstel74,Caldwell15}. When the surface is properly patterned, it can sustain also localised surface phonon polaritons,  confined in extremely sub-wavelength volumes and characterised by quality factors and Purcell enhancements unparalleled in plasmonic systems \cite{Schuller09,Caldwell13}. Patterning, apart from creating the localised modes, also acts as a grating for phase-matching to propagating surface polaritons \cite{Greffet02}, allowing to tune their dispersion, and making it possible to bring them in resonance with the localised ones. 
The possibility to couple long lived, optically accessible localised resonances to low-loss propagative modes, with propagation lengths of hundreds of micrometers \cite{Huber05}, hints to the tantalising prospect to realise phonon-polariton based quantum circuits, overcoming the main problems hampering the development of quantum plasmonic circuits \cite{deLeon12}.
Here, using a silicon carbide (SiC) surface patterned by micrometer-sized cylinders, we demonstrate tunable control of the surface polariton dispersion and we present clear evidence of spectral anticrossing between localised and surface modes, implying a coherent, reversible energy exchange between them \cite{Auffeves10,Konrad15}. Our work thus validates the different building blocks for a novel technological platform for optical and quantum mid-infrared applications. 

How tightly light of a given frequency may be confined is limited by the bandwidth of spatial frequencies available. The most famous example of this is the Abbe diffraction limit but the phenomenon is pervasive.  Piecewise homogeneous material systems can sustain electromagnetic resonances localised around interfaces where the permittivity changes sign, the out-of-plane wavevector becoming imaginary and the bandwidth of spatial frequencies in-plane increasing. For a flat surface between air and a material with negative permittitivity $\epsilon(\omega)$, this leads to surface modes characterised by the well known dispersion\\ 
\begin{equation}
\label{surfacedispersion}
q = \frac{\omega}{c} \sqrt{\frac{\epsilon\left(\omega\right)}{\epsilon\left(\omega\right)+1}}
\end{equation}
where $q$ is the in-plane wavevector and $c$ is the speed of light.
Surface plasmon polaritons are well known surface modes in metals, whose Drude permittivity $\epsilon_{\text{D}}(\omega)=1-\frac{\omega_{\text{P}}^2}{\omega^2}$ becomes negative due to the coupling with collective plasma excitations in the region below the plasma frequency $\omega_{\text{P}}$ \cite{MaierBook}.
Strong field localisations are achievable in plasmonic systems, with good applications in waveguiding \cite{Takahara98} and usually inefficient processes such as Raman spectroscopy \cite{Schlucker14}. Still, plasmons are inherently lossy \cite{Khurgin15}, the modal energy spending half cycles as electron kinetic energy, leading to a  dominant loss channel of electron-electron scattering occurring on the 0.01-ps scale \cite{Khurgin11}, thus making it challenging to integrate them in quantum technology architectures \cite{deLeon12,Tame13}.
As an alternative platform to metals, also polar dielectrics support surface polaritons in-between the frequencies of the transverse optical phonon, $\omega_{\text{TO}}$, and the longitudinal optical phonon, $\omega_{\text{LO}}$, where the Lorentz permittivity 
$\epsilon_{\text{L}}(\omega)=\frac{(\omega^2-\omega_{\text{LO}}^2)}{(\omega^2-\omega_{\text{TO}}^2)}$ becomes negative as a result of light coupling to oscillations of the ions. The damping of the ionic oscillations occurs on the 1-ps scale, two orders of magnitude slower than electron damping in metals. The resulting modes, called surface phonon polaritons \cite{Borstel74}, have been exploited for a number of applications, from enhanced energy transfer \cite{Shen09}, to thermal coherent infrared emission \cite{Greffet02}, superlensing \cite{Taubner06}, near field optics \cite{Taubner04}, and sensing \cite{Hillenbrand02}.
Analogously to localised plasmonic resonances, localised phonon resonances also appear in subwavelength dielectric systems. Mutschke \cite{Mutschke99} carried out explicit investigations into the infrared properties of small SiC particles of various polytype observing morphology dependent resonances analogous to particle plasmons. Subwavelength SiC whiskers have also been shown to support both localised electrostatic and propagative Fabry-Perot modes \cite{Schuller09}. Recently advances in fabrication procedures have allowed for the creation of user-defined cylindrical SiC nano-resonators on SiC substrate \cite{Caldwell13,Chen14}. The supported modes exhibit quality factors exceeding the theoretical limit for plasmonic resonators and Purcell factors 4 orders of magnitude higher than comparable plasmonic systems.

%%%%%% 
For our studies we used a $9.7\mu$m thick planar 3C-SiC substrate, on which subwavelength cylindrical resonators of height $\simeq800$nm and diameter $\simeq1\mu$m were fabricated by ICP RIE in square $70\times 70$ pillar arrays of varying periods from $5\mu$m to $7\mu$m. Full details on the fabrication process can be found in the methods section. The planar surface supports a surface mode whose dispersion, $\omega^{\text{s}}_{\vect{q}}$, is given by \Eq{surfacedispersion} with $\epsilon(\omega)$ the dispersive SiC dielectric function. While the surface polariton dispersion usually lies outside the lightcone, the periodic patterning of the substrate results in a period dependant band folding of the dispersion at the edge of the first Brillouin zone, as illustrated in Fig.~\ref{fig:Fig1}(a), thus making them optically accessible. In the inset the electric field norm of a surface mode is plotted.
The isolated cylinder-on-substrate system supports a number of modes as discussed in Ref. \cite{Chen14}. For the remainder of this Letter we will only consider the monopolar mode whose electric field norm is shown in the inset of Fig.~\ref{fig:Fig1}(b). This mode is a longitudinal mode of the cylinder mediated by the substrate resulting in charging of the pillars with neutrality assured by the interstitial substrate \cite{Caldwell13}. In the same inset we also show a SEM image of a single pillar.
The eigenmodes of the resonator array are periodic Bloch waves, with charges oscillating between the pillars and the substrate in between, whose dispersion $\omega_{\vect{q}}^{\text{m}}$ is derived in the methods section and illustrated in Fig.~\ref{fig:Fig1}(b) for a selection of periodicities. In Fig.~\ref{fig:Fig1}(c) a SEM image of the sample is shown.

\begin{figure}
\includegraphics[width=\linewidth]{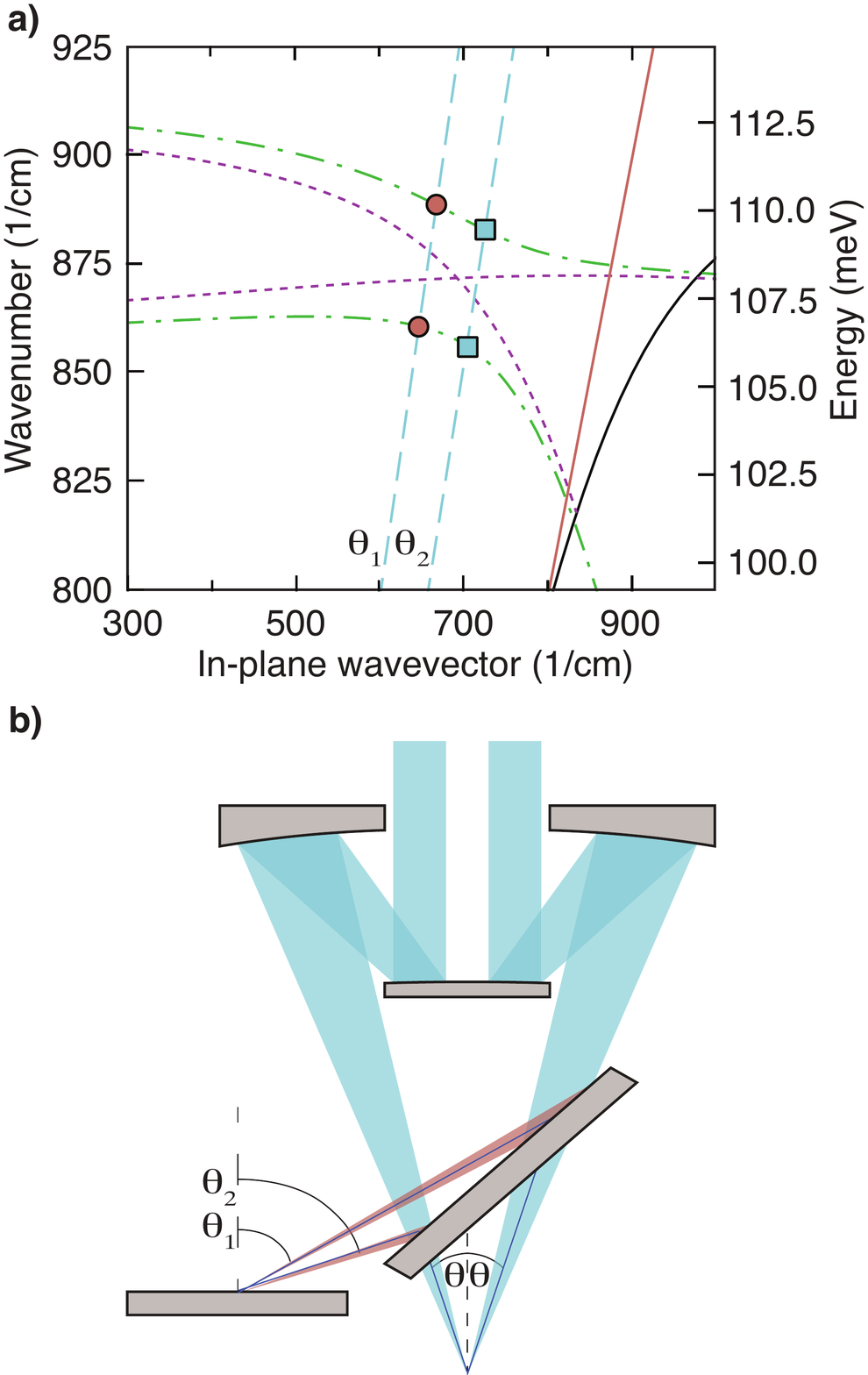}
\centering
\caption{\label{fig:Fig2} {\bf Theoretical model.} a) Dispersion for array period $6.25\mu$m and coupling constant $g_0 =1.63$meV ($13.1$/cm). Purple dotted lines are the constituent monopolar and folded surface phonon polariton branch, coupled normal modes are green dash-dotted lines. Blue dashed lines represent the two different angles $\theta_1$ and $\theta_2$ sampled by the dual illumination. b) Illustration of the function of the grazing incidence objective. A Schwarzschild light path is indicated by angularly symmetric blue rays. The mirror at the bottom rotates the symmetric rays, initially at angle $\theta$, giving the dual non-normal illumination at $\theta_1$ and $\theta_2$ indicated by red rays.
}\end{figure}

As the folded dispersion of the surface phonon polaritons eventually intersect the almost dispersionless localised monopolar mode, we expect the two modes to interact. Introducing a phenomenological $g_0$ Rabi frequency coupling the two, we can describe the coupled system, in the rotating wave approximation, by the Hamiltonian
\begin{equation}
\label{eq:Ham}
\mathcal{H} =  \sum_{\vect{q}} \hbar\omega_{\vect{q}}^{\text{m}} \hat{a}_{\vect{q}}^{\dagger} \hat{a}_{\vect{q}} + \hbar\omega_{\vect{q}}^{\text{s}} \hat{b}_{\vect{q}}^{\dagger} \hat{b}_{\vect{q}} 
+ \hbar g_0 \left(\hat{a}_{\vect{q}}^{\dagger} \hat{b}_{\vect{q}} + \hat{a}_{\vect{q}} \hat{b}_{\vect{q}}^{\dagger}\right)
\end{equation}
where $\hat{a}_{\vect{q}}$ and  $\hat{b}_{\vect{q}}$ are the bosonic creation  operators for the monopolar modes and surface modes respectively. As detailed in the methods section, the Hamiltonian in \Eq{eq:Ham} can be diagonalised in terms of two free normal modes, whose annihilation operators read
\begin{eqnarray}
\label{eq:Hop}
\hat{p}^+_{\vect{q}} &=& X_{\vect{q}} \hat{a}_{\vect{q}}+Y_{\vect{q}} \hat{b}_{\vect{q}}\\
\hat{p}^-_{\vect{q}} &=& Y_{\vect{q}} \hat{a}_{\vect{q}}-X_{\vect{q}} \hat{b}_{\vect{q}}
\end{eqnarray}
where $X_{\vect{q}}$ and $Y_{\vect{q}}$ are the Hopfield coefficients \cite{Hopfield58} describing the mixing of surface and localised modes and the frequency  of the normal modes is
\begin{equation}
\label{eq:PolDisp} 
\omega_{\vect{q}}^{\pm} = \frac{\omega_{\vect{q}}^{\text{m}} + \omega_{\vect{q}}^{\text{s}} \pm \sqrt{\left(\omega_{\vect{q}}^{\text{m}}-\omega_{\vect{q}}^{\text{s}}\right)^2 + 4 g_0^2}}{2}.
\end{equation}
The simulated electric field norm for a mode of the coupled system is shown in Fig.~\ref{fig:Fig1}(d), where for comparison, and in order to highlight the subwavelength character of the coupling, we explicitly show a typical wavelength for the resonant surface mode.
A typical dispersion of the normal modes, from \Eq{eq:PolDisp}, is shown by the dot-dashed lines in Fig.~\ref{fig:Fig2}(a), in which it is clear how the coupling between localised and surface modes leads to an anticrossing in the dispersion of the normal modes.
The fabricated arrays were measured by FTIR microscopy in reflectance mode utilising a grazing incidence objective. The objective illuminates directionally and the sample is aligned so the peak incident intensity is along the principal axis of the resonator array.
High-angle illumination is achieved by use of a mirror to rotate the incident beam onto the sample resulting in a dual peaked angular excitation as illustrated in Fig.~\ref{fig:Fig2}(b). This allows two slices of the polariton dispersion to be measured simultaneously as shown in Fig.~\ref{fig:Fig2}(a).
Note that we have until now neglected losses in our theoretical treatment, on account of the large quality factors of both localised and surface phonon polaritons. Still, it is important to remember that the anticrossing shown in Fig.~\ref{fig:Fig2}(a) is present only if the Rabi frequency $g_0$ is larger than the losses of both modes, including pure dephasing \cite{Auffeves10}, a condition usually referred to as strong coupling regime. Observing an anticrossing in the system spectrum thus unequivocally proves that a quantum of energy is reversibly and coherently exchanged between the two modes, fulfilling  the main requirements to use them as building blocks for quantum circuit architectures \cite{Konrad15}.
\begin{figure}
\centering
\includegraphics[width=10.0cm]{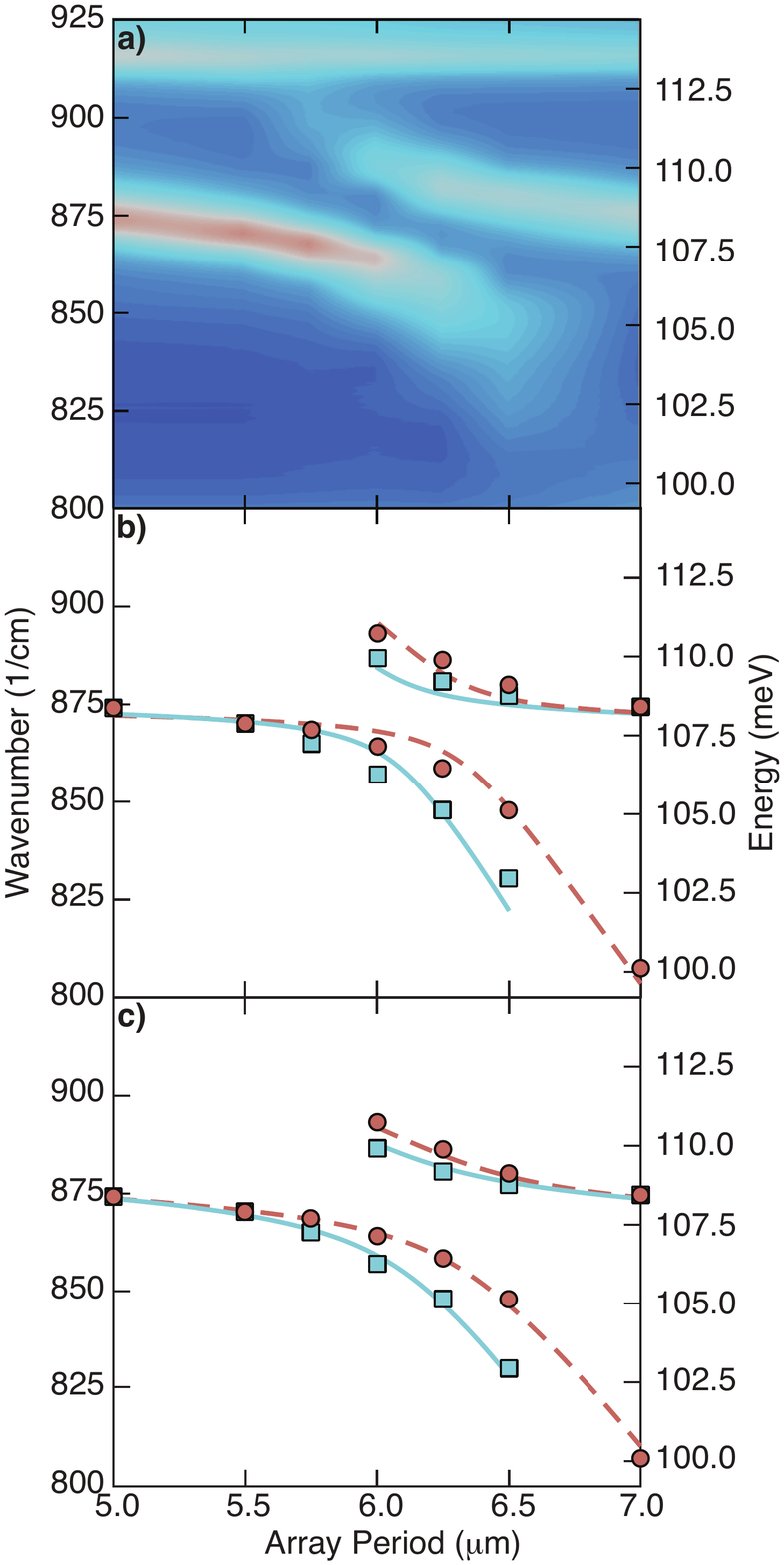}
\caption{\label{fig:Fig3} {\bf Strongly coupled normal modes.} The top panel (a) shows the
experimental reflectance map of SiC cylinder arrays of varying period. The almost dispersionless mode at $113.75$meV ($917.4$/cm) is the transverse dipole resonance discussed elsewhere \cite{Caldwell13}. The peaks extracted from the reflectance map are given in the lower panels for the larger angle by blue squares and the smaller by red circles. Solid blue lines and dashed red lines are the corresponding fits, enacted using book values for the dielectric constants of SiC (b) \cite{Pitman08}, or fitting also the dielectric constants of SiC (c).}
\end{figure}
The experimental reflectance map we obtained is given in Fig.~\ref{fig:Fig3}(a) as a function of the array period, that is tuning the surface mode resonance. The data exhibits a clear anticrossing when the two modes are resonant, showing that the system is indeed in the strong coupling regime.
Peak positions were then extracted from the experimental reflectance map, clearly highlighting the presence of peaks from two different angles, not apparent in Fig.~\ref{fig:Fig3}(a) due to the small angular shift and finite linewidth.
 The data was then fitted, following the procedure detailed in the methods section, to the normal mode dispersion given in Eq.~\ref{eq:PolDisp}. The peak positions extracted from the experimental data and the resulting fits are given  in Fig.~\ref{fig:Fig3}(b), where we explicitly show the dispersions at the two different angles. 
The model reproduces well the anticrossing, within errors of the order of $~1$ meV ($8$/cm), a similar magnitude to those reported in previous simulations using finite element modelling \cite{Caldwell13}. Those errors have been attributed to modifications of the dielectric properties of SiC near the surface due to the strain induced in SiC grown on Si substrates due to the mismatch of lattice parameters \cite{Ferro06}.
In order to verify this hypothesis we repeated the fitting procedure using the high and low frequency values of the dielectric constants and the TO phonon frequency as additional fit parameters. The resulting values for the dielectric parameters differ less than $5\%$ from the standard values found in the literature \cite{Pitman08} and the TO phonon shifts just $0.74$meV ($6.1$/cm), but they lead to a dramatic improvement of the fits, shown in Fig.~\ref{fig:Fig3}(c).
The maximal value of the fitted Rabi frequency is $g_0=2.55$meV ($20.6$/cm), leading to a ratio between $g_0$ and the bare frequency of the excitation of the order of $10^{-2}$, thus justifying a posteriori the rotating wave approximation we used in \Eq{eq:Ham} \cite{Anappara09}.
We also extracted the linewidths of the different normal modes from the reflectance map.  
In order to gain some understanding of their behaviour, we fitted them assuming they are sums of the constituents' ones, weighted by the relative Hopfield coefficients \cite{DeLiberato14}
\begin{eqnarray}
\label{eq:LW}
\gamma^{+}_{\vect{q}} &=&  \gamma^{\text{m}} \lvert X_{\vect{q}}\lvert^2+\gamma^{\text{s}}_{\vect{q}}\ \lvert Y_{\vect{q}}\lvert^2\\
\gamma^{-}_{\vect{q}} &=& \gamma^{\text{m}} \lvert Y_{\vect{q}}\lvert^2 + \,\gamma^{\text{s}}_{\vect{q}}\ \lvert X_{\vect{q}}\lvert^2
\end{eqnarray}
where $\gamma^{\text{m}}$ is the linewidth of the monopolar mode, essentially constant over the measured region, and $\gamma^{\text{s}}_{\vect{q}}$ is the dispersive linewidth of the surface mode. Experimental data and the fits are given in Fig.~\ref{fig:Fig4}.
Notwithstanding the simplicity of the model, that neglects both the broadening due to the steep dispersion of the surface mode, and the dispersive photonic losses due to the presence of the pillars, we still obtain a good numerical agreement, with errors below $0.5$ meV ($4$/cm). 
Notice that the linewidth at the anticrossing is of the order of $1$ meV ($8$/cm). In the densest array we considered, energy is thus coherently transferred between monopolar and surface modes roughly $4$ times before escaping.
\begin{figure}
\includegraphics[width=7cm]{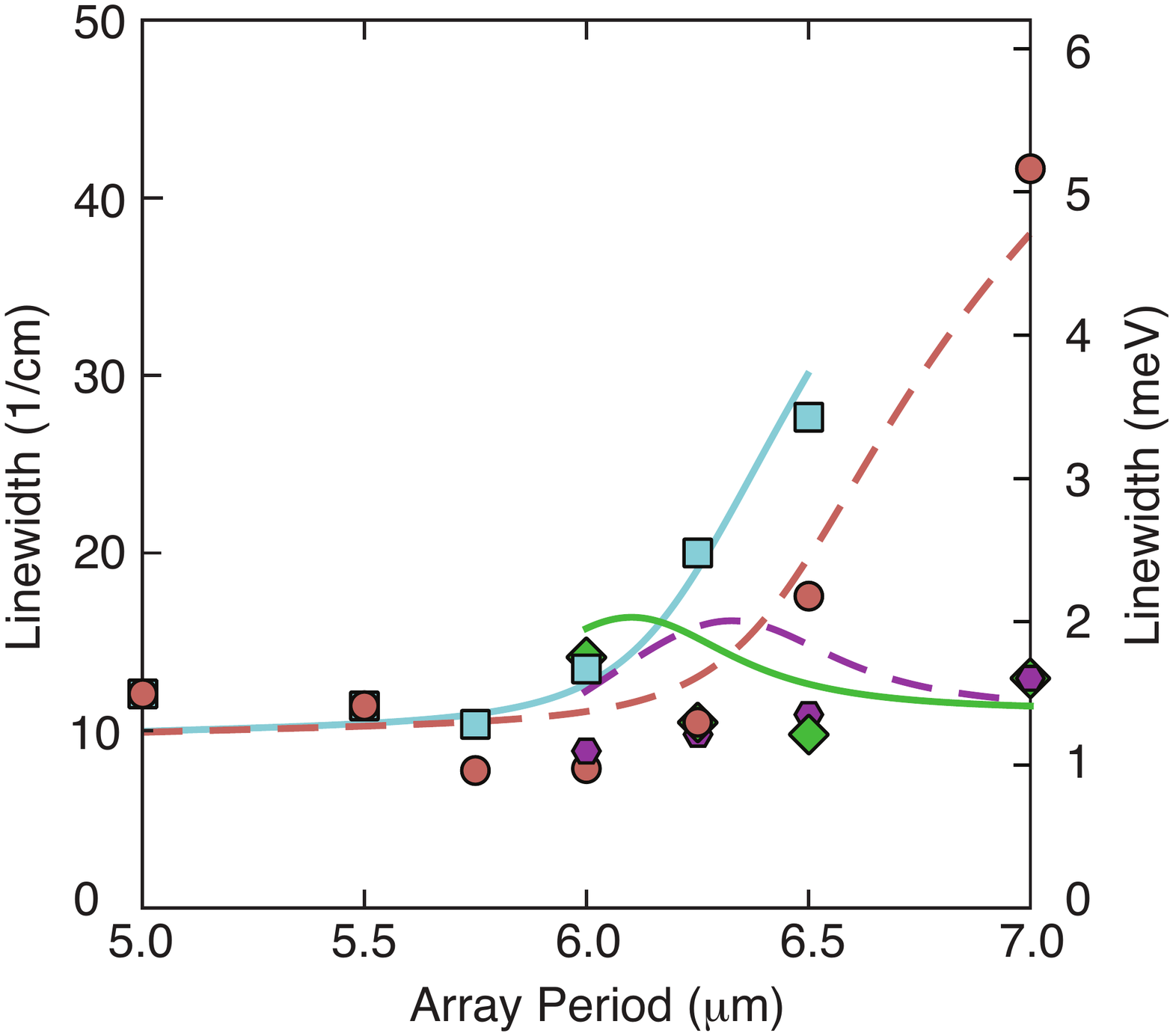}
\centering
\vspace{-2cm}
\caption{\label{fig:Fig4} {\bf Linewidths of the normal modes.} Experimental linewidths of the lower (upper) normal mode for the smaller angle data are given by red circles (purple hexagons), and the respective fits are the red (purple) dashed line. Blue squares (green diamonds) and blue (purple) solid lines are instead used for the larger angle lower (upper) polariton data.}
\end{figure}

% Conclusions.
In summary, we have demonstrated the tunability of surface phonon polaritons dispersion varying the periodicity of the surface patterning. This allowed us to observe a clear spectral anticrossing between localised and surface phonon polaritons, proving that coherent, reversible energy exchange is possible between them.

In combination with the high confinements and Purcell enhancements recently observed in user-defined structures, the present Letter takes a decisive step in demonstrating the versatility and tunability of phonon polaritons for optical and quantum applications in the mid-infrared spectral region. In particular, the coherent interplay between localised and propagative, nonradiative modes, together with the relatively large quality factors achievable in those systems, could make it possible to design quantum architectures similar to quantum plasmonic circuits, but without many of the limitations due to plasmon intrinsic losses.

%\appendix{}
\section{Methods}
\textbf{Fabrication:} The surface phonon polariton resonators were fabricated starting from a polished $9.7\mu$m thick layer of $<100>$ oriented 3C-SiC,  grown heteroepitaxially on a $<100>$ Si substrate (NOVASIC). Patterning was carried out via standard liftoff process, using a bilayer of MMA-PMMA (thickness $250$nm and $150$nm respectively) exposed by electron beam lithography. The bilayer allowed a deposition of $120$nm of Ni hardmask via electron beam evaporation. The sample was kept in acetone until the unpatterned Ni was completely removed and subsequently dry etched via ICP RIE in a SF6 and Ar chemistry at $0.7$mTorr pressure, $280$W of bias power and $800$W of ICP power. Finally, the Ni hardmask was removed with a fuming nitric acid wet etch for $20'$. The dry etch was calibrated to obtain $811\pm 8$nm high structures and shows an etching angle of $86.5$ degrees.\\

\textbf{Measurements:} The reflectance map in Fig.~\ref{fig:Fig3}(a) was recorded using a Bruker Hyperion 2000 FTIR microscope in reflectance mode. Data was recorded with use of the grazing incidence objective which utilises a mirror to rotate the incident light cone onto the sample resulting in a double peak in the angular intensity spectrum as illustrated in Fig.~\ref{fig:Fig2}(b).\\

\textbf{Theoretical Modelling:} The dispersion of the uncoupled cylinder array is described by a tight binding model as 
\begin{equation}
\omega^m_{\vect{q}} = \omega^c\left(\frac{1+\sum_{n\neq 0} e^{-i n q R} \beta_n}{1+\Delta\alpha+\sum_{n\neq 0} e^{-i n q R} \alpha_n}\right)
\end{equation}
where $n$ indicates a discrete resonator, $\alpha_n$, $\beta_n$ and $\Delta\alpha$ are overlap integrals as defined in \cite{Yariv99}, R is the array period, and $\omega^c$ is the frequency of the monopolar mode of a single cylinder. We consider only nearest neighbour interactions in the relevant limit where the coupling between resonators is small due to the large separations, and the tight-binding equation simplifies to
\begin{equation}
\omega^m_{\vect{q}} = \omega^c \left(1 - \frac{\Delta \alpha}{2} + \kappa_1 \cos{\left(q R\right)}\right)
\end{equation}
where $\kappa_1 = \beta_1 - \alpha_1$. The remaining tight-binding parameters $\kappa_1$ and $\Delta\alpha$ are assumed to have a dipole-dipole like dependancy on the resonator separation, modelled as $1/R^3$ . The appropriateness of this model was independently confirmed by numerical simulations carried out in the RF module of COMSOL Multiphysics.
As the surface mode wavelength is larger than the pillars separation, the surface-monopole coupling strength $g_0$ is assumed to vary superradiantly with the in-plane resonator density $\rho$ as $\sqrt{\rho}$ when the array period is varied. This is analogous to previously reported scalings in systems consisting of surface plasmons interacting with molecular excitons, where in our case the resonators act as effective molecules \cite{Cade09}.

The Hamiltonian in \Eq{eq:Ham} can be put in diagonal form by diagonalising the corresponding Hopfield-Bogoliubov matrix for each value of the in-plane wavevector $\vect{q}$
\begin{eqnarray}
H_{\vect{q}}&=&
\left[
\begin{array}{cc}
 \omega^{\text{m}}_{\vect{q}} &   g_0  \\
g_0  & \omega^{\text{s}}_{\vect{q}}      
\end{array}
\right]
\end{eqnarray}
whose eigenvalues are the frequencies in \Eq{eq:PolDisp} and the respective eigenvectors $\lbrack X_{\vect{q}},Y_{\vect{q}}\rbrack$ give the Hopfield coefficients appearing in \Eq{eq:Hop}.

The surface phonon polariton mode results from strong coupling of photons to the transverse phonon resonances of the crystal. It can be described \cite{Todorov14} in the same Hopfield-Bogoliubov framework we used previously as a linear superposition of free photons  
and optical phonons components, whose Hopfield coefficients we denote $C_{\vect{q}}$ and $D_{\vect{q}}$ respectively. As the free photon dispersion is very steep those coefficients depend strongly on the wavevector. The linewidth of the surface mode therefore also obeys the equivalent of \Eq{eq:LW},
\begin{equation}
\label{eq:LW2}
\gamma^{\text{s}}_{\vect{q}} = \gamma^{\text{ph}} \left|C_{\vect{q}}\right|^2 + \gamma^{\text{TO}} \left|D_{\vect{q}}\right|^2
\end{equation}
where $\gamma^{\text{ph}}$ and $\gamma^{\text{TO}}$ are respectively the loss rates of the photonic and phononic components. \\

\textbf{Fitting:} In order to fit the spectrum of the system as a function of the array period, as shown in Fig.~\ref{fig:Fig3}(b),
least squares fits were carried out for the tight binding parameters $\kappa_1, \Delta\alpha, \omega^c$, the two incident angles $\theta_1, \theta_2$ shown in Fig.~\ref{fig:Fig2}(a), and the parameter $\zeta$, linked to the coupling strength as $g_0=\zeta\sqrt{\rho}$. For Fig.~\ref{fig:Fig3}(c) also the high and low frequency dielectric constants $\epsilon_0$ and $\epsilon_{\infty}$ and the TO phonon frequency $\omega_{\text{TO}}$ were used as fitting parameters. 
The fitting procedure yielded $\theta_1=\left(48.54\pm 0.05\right)^{\circ}$, $\theta_2=\left(55.08\pm 0.08\right)^{\circ}$, $\epsilon_0=9.26 \pm 0.22$, $\epsilon_{\infty}=6.68 \pm 0.17$ and $\omega_{TO} = 97.85\pm0.07$meV  ($789.3 \pm 0.6$/cm).
To determine the linewidths in Fig.~\ref{fig:Fig4} we used the Hopfield coefficients fixed by the previous procedure, and then fitted the measured linewidths with \Eq{eq:LW} and \Eq{eq:LW2}, using $\gamma^{\text{TO}}$, $\gamma^{\text{ph}}$, and $\gamma^{\text{m}}$ as parameters.

\section{Acknowledgements}
S.A.M. acknowledges support from EPSRC programme grants EP/L024926/1 and EP/M013812/1, plus ONR Global.
S.D.L. is Royal Society Research Fellow and he acknowledges support from EPSRC grant EP/M003183/1.

\end{document}